\def\arcsec{$^{\prime\prime}$}
\def\arcmin{$^\prime$}
\def\deg{$^\circ$}
\def\apj {ApJ}
\def\apjl {ApJL}
\def\aj {AJ}
\def\mnras {MNRAS}
\def\aap {A\&A}
\def\nat {Nat}
\def\araa {ARAA}
\def\procspie {SPIE}
\def\pasj {PASJ}
\def\pasp {PASP}

\documentclass{iaus}
\usepackage{rotating}
\usepackage{graphicx}

\title [Imaging science and stellar populations: a short review] 
{Stellar populations -- the next ten years}

\author[Bland-Hawthorn]
{J. Bland-Hawthorn}

\affiliation{Anglo-Australian Observatory, PO Box 296, Epping, NSW 2121, Australia}

\pubyear{2007}
\volume{241}
\pagerange{119--126}
\date{January 2007}
\jname{Stellar Populations as Building Blocks of Galaxies}
\editors{A. Vazdekis \& R. Peletier}
\begin{document}

\maketitle

\begin{abstract}
The study of stellar populations is a discipline that is highly
dependent on both imaging and spectroscopy. I discuss techniques
in different regimes of resolving power: broadband imaging (R$\sim$4),
intermediate band imaging (R$\sim$16, 64), narrowband spectral
imaging (R$\sim$256, 1024, 4096). In recent years, we have seen
major advances in broadband all-sky surveys that are set to continue
across optical and IR bands, with the added benefit of the time
domain, higher sensitivity, and improved photometric accuracy.
Tunable filters and integral field spectrographs are poised to make
further inroads into intermediate and narrowband imaging studies
of stellar populations.  Further advances will come from AO-assisted
imaging and imaging spectroscopy, although photometric accuracy
will be challenging. Integral field spectroscopy will continue to
have a major impact on future stellar population studies, extending
into the near infrared once the OH suppression problem is finally
resolved.  A sky rendered dark will allow a host of new ideas to
be explored, and old ideas to be revisited.


\end{abstract}


\bigskip\noindent
{\bf\large 1. Introduction}

This conference has provided us with a timely reminder of why stellar
population studies continue to generate great interest in our quest
to understand stellar/galactic formation and evolution.  In a recent
summary, we reviewed the case for resolved stellar population studies
(Bland-Hawthorn \& Freeman 2006), and so refrain from repeating
these arguments here; another perspective is offered by Wyse
\& Gilmore (2006). The importance of near field studies has been
stressed most recently by Springel, Frenk \& White (2006):

\smallskip
\begin{quote}
At present, the strongest challenge to $\Lambda$CDM arises
not from the large-scale structure, but from the small-scale structure
within individual galaxies. It is a real possibility that the model
could be falsified by measurements of the distribution and kinematics
of matter within galaxies...
\end{quote}
\smallskip

\smallskip\noindent
The detailed study of the stellar content of galaxies is well
grounded in cosmology.  The field exploits many observational
techniques across almost the full electromagnetic spectrum. The
observations are so rich in detail that they present both theorists
and simulators with a strong challenge, i.e. to establish useful
empirical models that provide a context for these data, and identify
those observations that are of primary importance. It is no
exaggeration to say that contemporary data far outstrip theory and
simulation: our understanding of the observations is very rudimentary
when we consider even simplified toy models (e.g. closed vs. open
box), and especially so within the context of the $\Lambda$CDM
paradigm.

But here I concentrate on the observational techniques and address
in part how these are motivated by the scientific questions.  Given
that we are already overwhelmed by vast imaging data sets, is there
a case for considering new approaches to data collection?  In short,
the answer must be yes. More sophisticated techniques can yield
better data that in turn shed light on the existing data sets, and
may even simplify the paradigm rather than complicate it further.
It is often the case that more and better data can reveal weak
effects that are of paramount importance to our basic understanding,
a fact well understood by medical science and an issue that we
return to below.

Much like mathematical physicists who utilize the properties of
discrete or continuous functions, a stellar populations researcher
has the option to study discrete (point source photometry) or
continuous (diffuse light / surface photometry) data. Nowadays,
this distinction is somewhat artificial in that both frequently
arise in the same observations; often one is removed to get to the
other, e.g. surface brightness fluctuations (SBF) and the removal
of point sources.  It is worth noting that studies of resolved
stellar populations in excellent seeing are now achieving effective
surface brightness levels that are well below those of surface
photometry (Brown et al 2003; Bland-Hawthorn et al 2005, see
Appendix).

Imaging techniques, in particular imaging spectroscopy, will continue
to develop in the years ahead.  The review focuses on resolved
studies of galaxies, rather than multi-object astrometry and
spectroscopy (R $>$ 5000) of Local Group galaxies that I will defer
to a separate review.  I start with a reminder of the power of
imaging science (including imaging spectroscopy) before addressing
how it is likely to evolve in the years ahead. I address new
technologies and new concepts that may play an important role in
future stellar population studies, before suggesting new avenues
for future research.


\bigskip\noindent
{\bf\large 2. Imaging science -- what is it good for?}
  
This meeting has provided an opportunity to review what are the
essential characteristics of data that advance our understanding
of stellar populations.  A familiar proverb is that a picture is
worth a thousand words.  Our images captivate the imagination of
the general public, and astronomers have exploited this to great
effect in recent years, as witnessed by ongoing public support for
the Hubble Space Telescope (HST).  But astronomical machines are
sold on the scientific return, and from these images we learn about
stellar structures and morphologies, subcomponents (e.g. bulges),
substructures (e.g.  streams), small-scale vs. large-scale power,
and asymmetries, gradients and profiles.

The emergence of imaging spectroscopy (IFU, tunable filter) is a
result of the community wanting the photometric integrity of direct
imaging, with the benefits of spectroscopic information across the
field. Imaging spectroscopy provides significant advantages over
conventional slit techniques, and are less prone to losses and
artefacts due to seeing (Cecil 2006). We explore this topic in more
detail below, and conclude that imaging spectroscopy will become
more firmly entrenched in the next decade.

Since the 1960s, we have mastered the use of broadband images to
obtain limited information on stellar temperatures, luminosity,
surface gravity, mass, ages, dust content and star formation history
(for a recent review, see Bessell 2005).  What is generally true
is that more spectral bands over a broader spectral baseline provide
more information of the sources under study\footnote{This assertion
can be difficult to argue if the sensitivities in the narrower bands
lead to marginal detections, e.g. continuum sources vs. emission-line
sources.}.  Thus, we make use of intermediate (e.g.
Stromgren filters) and narrow spectral bands
to provide more specific information, in particular, better spectral
classification (e.g. Geneva-Copenhagen survey: Nordstrom et al 2004) 
or galaxy typing (e.g. COMBO-17 survey: Wolf et al 2003).

What is clear is the continued importance of the broadest possible
spectral baseline in stellar population studies.  The Balmer jump
($u$ band) region provides temperatures of OB stars and surface
gravities of cool stars, and metallicity information for stellar
type F and later. In combination with the $u$ band, a $b$ or $g$
filter provides metallicity information for stellar type A and later.
Filter bands spaced further apart provide temperature information
that is moderately independent of metallicity and gravity.  Accurate
$K$ band stellar photometry, in combination with optical bands, can
be used to unravel the age-metallicity degeneracy, and provide more
precise corrections for stellar extinction.


\bigskip\noindent
{\bf\large 3. Photometric precision}

A central tenet of the applied sciences that simply measuring the
same old quantities to increasing levels of precision will often
yield fundamentally new physical insight\footnote{Dyson (1999) has
argued that modern science began in 1729 with Bradley's astrometric
measurements that were recorded with an accuracy of 6 significant
figures. Bradley first established that the Copernican view was
correct, detected stellar aberration due to the Earth's motion, and
measured the speed of light to an accuracy of 1\%. In the next
century, Michelson spent several decades measuring the higher order
aberration coefficients with ever-increasing precision, with profound
consequences for physics.}. A good discussion of what is required
to achieve this in astronomical photometry is given by Stubbs \&
Tonry (2006).  The power of photometric precision is well demonstrated
by the Sloan Digital Sky Survey (Gunn et al 2006) that achieved
5$\sigma$ depths of ($u, g, r, i, z$) = (22.0, 22.2, 22.2, 21.3,
20.5). The SDSS is comparable in depth and sampling to the all-sky
surveys of the UK Schmidt (UKST) and Oschin Schmidt telescopes
during the 1980s.  However, the SDSS magnitude and colour precision
is unrivalled at ($r, u-g, g-r, r-i, i-z$) = (2\%, 3\%, 2\%, 2\%,
3\%), a fact that has led to an extraordinary bounty of new results.
A similar comparison can be made between the space-borne Galex UV
(Martin et al 2003) and the Ultraviolet Imaging Telescope surveys.

Arguably, precision photometry has provided the greatest impetus
to the study of stellar populations in recent years. In resolved
stellar imaging, examples include the use of non-parametric methods
to derive believable star formation histories in nearby dwarf
galaxies (Aparicio, this meeting) and in Local Group galaxies (Brown
et al 2006).  In photometric imaging, some of the most
impressive uses of precision photometry involve distance determinations
for galaxies: e.g. cepheid distances (Freedman et al 2001); surface
brightness distances to galaxies (Tonry et al 2001) and its close
cousin, "tip of the red giant branch" distance estimates (Tully et
al 2006).  At much higher redshift, other uses include redshift
determination through the photometric drop-out technique (Steidel
et al 1996).  All of these techniques exploit known properties of
the stellar populations.

If our goal is to learn more about a source by dividing up the
spectral domain into more bands, this requires a higher level of
photometric precision within and between the bands. This is because
the errors combine in quadrature and quickly propagate through the
analysis when one wants to compare complex intercombinations of
photometric colours.

Pushing to even higher levels of precision, there is a strong case
to be made for sub-millimag photometry of hundreds of thousands of
stars in order to establish accurate relative ages through
asteroseismology. This was the primary motivation for the ESA
Eddington satellite which is unfunded at the present time.
Science in the time domain is broached in the following section.


\bigskip\noindent
{\bf\large 4. Imaging science -- how will it evolve?}

\smallskip\noindent
{\bf Time domain.}  
The time domain is clearly an area of parameter space that has been
barely explored to date. The MACHO project (Alcock et al 1997) taught
us that there is much to be learnt from stellar populations
through variability studies. 

Two important survey projects, Pan-STARRS\footnote{A related project
in the southern hemisphere is SkyMapper, essentially a single
Pan-STARRS telescope at Siding Spring Observatory.} (north) and the
Large Synoptic Survey Telescope (south), are set to dominate the
optical landscape in the next decade by extending beyond SDSS both
in the time domain and in source sensitivity (combined exposures).
The first of these will make use of four 1.8m telescopes, each with
a 3\deg\ field.  Pan-STARRS will repeatedly survey the sky with
30-60 sec exposures and achieve a single-exposure depth of 24.0 AB
mag (5$\sigma$). The LSST is a single 8.4m telescope with a 3.3\deg\
field that will survey the sky in 15 sec exposures, with a
single-exposure depth of 24.0 AB mag (6.5$\sigma$). Both surveys
will make use of $g, r, i, z$ filters with an additional red $y$
filter for the LSST. The survey \'etendue (area $\times$ solid
angle, $A\Omega$) of Pan-STARRS and LSST is 60 and 190 respectively.

Both LSST and Pan-STARRS will reveal an extraordinary richness of
new data across the sky, inter alia, identifying homogeneous stellar
populations (miras, cepheids, etc.) by their distinct temporal
behaviour, proper motions, colours, etc. These populations in turn
will provide age and metallicity information as a function of
position throughout Local Group galaxies (Luck et al 2006). One of
the enduring legacies of these surveys will be the prospect of
selecting clean stellar populations untainted by unresolved binaries;
the binary fraction can easily exceed 25\% in a typical population.

\smallskip\noindent
{\bf New deep wide-field surveys.} The upcoming UKIRT Infrared Deep
Sky Survey (UKIDSS; north) and Visible and Infrared Survey
Telescope for Astronomy (VISTA; south) both utilize IR-optimized
4m telescopes in order to provide deep $ZYJHK_S$ band photometry
over large swaths of the sky.  The UKIDSS Large Area Survey will
reach 18.4 mag (5$\sigma$) for stellar sources over a 4000 deg$^2$
field, and 0.5 mag deeper for 3200 deg$^2$ in galactic fields.
This is to be compared with the 2MASS and DENIS all-sky surveys
that reached 14.2 and 14.0 mag respectively. But first off the 
ranks will be the VLT Survey Telescope (VST 2.6m) which becomes 
operational in 2007. The OmegaCAM imager has a 1\deg\ field and
will supply good-seeing SDSS-style data over a 4500 deg$^2$ field.

Other noteworthy systems are the IMACS imager on the Magellan 6.5m
telescope, and the SuprimeCam imager on the Subaru 8.2m telescope.
SuprimeCam sits at the f/2 focus and has an $A\Omega$ product of
about 9 (Miyazaki et al 2002). Remarkably, engineers at Canon Inc.
believe that the \'etendue can be increased by an order of magnitude
$-$ the so-called HyperSuprimeCam concept $-$ which would have
comparable survey power to Pan-STARRS. But the primary science
driver for the Subaru cameras is deep multiband optical exposures
for observational cosmology, and these share the telescope with a
host of other instruments, unlike the dedicated imagers on Pan-STARRS.

\smallskip\noindent
{\bf Intermediate band all-sky surveys.}  In an era of precision
photometry, it would seem an appropriate time to revisit the power
of intermediate band photometry for identifying specific populations.
This was to be a mainstay of the ESA GAIA satellite mission in the
next decade. But in its latest incarnation, the intermediate
photometric filters will not be employed. This is unfortunate as
we will still need precise photometry for the billion or more sources
to be targetted by this important astrometric mission (Perryman et
al 2001; Wilkinson et al 2005).  It is clear however that the 5$-$6D
phase space information will have a huge impact on our understanding.
The mission will allow for millions of stars to be characterised
by their distance, spectral type, luminosity and so forth.  It is
interesting to observe the scientific impact of even relatively
crude photometric distances to stars (e.g. SDSS tomography of the
Galaxy; Juric et al 2005).

\smallskip\noindent
{\bf Short-term gains: fourfold increase in $A\Omega$.}
Conventionally, we build large imaging mosaics to work in
fast beams in order to enhance the \'etendue of an imaging survey.
The unvignetted field however is typically larger but not fully
exploited to minimize the cost of detector real estate. For future
wide-field instruments, it may be interesting to ask whether full
Nyquist sampling is required over the field, particularly with a
view to the dithering requirement in most applications today.
If the pointing and guiding models are robust, can we simply match
the median seeing to the pixel scale, and attempt to double the
available field of view? If we dither with sub-pixel offsets, can
we recover the psf (e.g. Fruchter \& Hook 2002)? This is an experiment
that is worth exploring.

\smallskip\noindent
{\bf Short-term gains: fourfold increase in resolution and sampling.}
Driver and colleagues have amply demonstrated the advantage of even
moderate improvements over the SDSS in the median seeing.  They
used the Wide Field Camera on the INT 2.5m to image 10,000 galaxies
over a 37 deg$^2$ field complete to $B\sim$ 24 mag. Their
median seeing is roughly a factor of 2 better than the SDSS (Lemon
et al 2002) resulting in markedly better bulge-disk decompositions
(Allen et al 2006). This emphasizes the importance of imaging cameras
on good sites for resolved studies.  My colleagues and I have
recently completed more than 40h of optical imaging on nearby
galaxies using the GMOS imaging spectrograph on Gemini South with
seeing in the range $0.55-0.65$\arcsec\ over the full 5.5\arcmin\
field. Two hour exposures result in 3$\sigma$ point source detections
of 27.0 AB mag (Bland-Hawthorn et al 2005). Few galaxies have
been mapped to this level of sensitivity. Important exceptions
are the few nearby galaxies that have been studied with the HST,
as we have seen at this meeting. 

\smallskip\noindent
{\bf Sub-arcsecond imaging over the widest possible field.} We have
barely begun to exploit wide-field imagers on the best observing
sites to study nearby galaxies.  The LSST, Pan-STARRS and VISTA
surveys are expected to achieve $\approx$0.7\arcsec\ FWHM median
image quality, although a smaller subset of observations at lower
sensitivity may achieve 0.5\arcsec\ FWHM.

Consistently higher quality data will require dedicated photometric
imagers. Both Magellan IMACS and Subaru SuprimeCam have recorded
$<$0.5\arcsec\ psf FWHM in optical bands over their half-degree
fields.  A superior system is the HyperSuprimeCam facility now under
discussion at Subaru: there is the prospect of achieving a 2\deg\
corrected field at the prime focus of this 8m telescope! Facilities
like these are needed to observe of order 10,000$-$100,000 galaxies
to examine the environmental influences on galaxies, and more
specifically stellar populations within individual components.

Beyond the well established morphology/luminosity dependencies on
environment, Blanton et al (2006) and Park et al (2006) assert that
the environmental influences on SDSS galaxies are very weak.  But
it is likely that higher resolution imaging of these large galaxy
samples will reveal stronger influences once the different galactic
components are properly separated.

Stellar population studies will always benefit from improved
image quality.  When studying the resolved stellar populations, the
problem remains even at 0.5\arcsec\ FWHM of distinguishing stars
from background galaxies, particularly in the diffuse outer parts
of nearby galaxies (see Ellis \& Bland-Hawthorn (2007) for a detailed
analysis of this point $-$ the {\tt GalaxyCount} java calculator
is freely available at www.aao.gov.au/astro/GalaxyCount).  This
ambiguity all but disappears in HST studies.  Clearly, there is a
strong case for a long-term HST program to target a large sample
($\sim100$) of nearby galaxies in a magnitude-limited survey.

\smallskip\noindent
{\bf Near diffraction-limited imaging over the widest possible field.} 
There are numerous advantages to be gained from sampling at higher
angular resolution, as evidenced by the extraordinary gains of HST
with its small telescope aperture.  The most important gains include:
lower sky background per pixel, sources observed at higher intrinsic
resolution, better star/galaxy and star/star separation. These are
some of the reasons behind the major investment in adaptive optics
and the projected enormous investment in ELTs.

Beyond Hubble, what prospects are there to achieve $\le$0.25\arcsec\
seeing at optical/IR wavelengths with a wide-field ground-based
facility?  At optical wavelengths, the only prospect under discussion
is the 2m PILOT survey proposed for Dome C in Antarctica (Burton
et al 2005). There has been talk of "quasi" HST-quality imaging
from simple ground-layer correction, although it remains to be seen
whether a 2m telescope located at 30m above the snowpack can achieve
this performance for a useful fraction of the time. The Australian
community is currently engaged in a study to look at the practicality
of this project with two commercial telescope builders. Such a
facility could be devoted to the long-term study of stellar populations
in nearby galaxies, and indeed is one of the key science drivers
for the project.

An interesting case has been made for ``Lucky imaging" whereby $I$
band images are read out at high frame rates using a low-noise E2V
Technologies L3CCD (Law, Mackay \& Baldwin 2006). In the best
conditions, they achieve diffraction limited images over roughly
an arcminute using the NOT 2.6m, but the overall operational
efficiency is low.

At IR wavelengths, things look more promising.  Existing AO systems
have begun to deliver on science (see cfao.ucolick.org/links for
links to these projects).  The Keck AO system has produced more
than 100 research papers over the past decade (Liu 2006). Typical
Strehl ratios at K are 0.2 with a best operational value of 0.4.
Both Gemini and the VLT are close to commissioning multi-conjugate
AO (MCAO) systems that promise 0.2\arcsec\ image quality on 1-2\arcmin\
scales.  The Gemini-N Altair system combined with the near IR imager
NIRI has been used to study stellar populations in M31 (Olsen et
al 2006). The system can achieve Strehl ratios of about 20-25\% at
H and K, with psf variations $\sim$3\% and photometric accuracies
of $\sim$5\% (Olsen 2006, private communication), both with natural
and laser guide stars. Soon, the MCAO f/30 focus will be available
at Gemini South and will produce a corrected field of 1\arcmin, and
a usable field of up to 2\arcmin. This focus will feed instruments
such as the GSAOI infrared imager and the Flamingos-2 infrared
spectrograph. Comparable performance is offered by the VLT/NACO
system as demonstrated by Cresci et al (2006) in a study of
intermediate-redshift galaxies in 20 fields close to natural guide
stars.

Liu (2006) provides a good discussion of the advantages of ground-based
AO over HST in the near IR.  These include the use of novel
instruments, higher observing efficiency (although terrestrial
weather is more of a hindrance than space weather), and 3-4$\times$
better spatial resolution in the near IR (0.05\arcsec\ fully
corrected).  In contrast, the disadvantages are the need for tip-tilt
stars, a fully corrected field on sub-arcminute scales, and complex
and variable psf leading to heterogenous data.

But do we need to achieve the diffraction limit for population
studies? This is not altogether clear. In my opening remarks, I
alluded to the importance of studying stellar populations out to
Virgo. It would seem that stellar crowding demands the smallest
possible psf. The diffraction limit is really only achieved in
practice at high Strehl ratios, i.e. where the 80\% EE diameter is
close to the angular diffraction limit.  Existing ground-based
studies suggest that this is going to be very difficult to achieve
in J and H, and challenging at K. However, in an important study,
Olsen, Blum \& Rigault (2003) find that high Strehl ratios may not
be required in crowded fields because of the SBF contribution of
unresolved stars.

So are there compelling science drivers for rigorously diffraction
limited imaging (extreme AO), beyond that of identifying material
and planets around nearby stars?  A strong case can be made for
studies of the diffuse stellar populations in the vicinity of compact
sources (e.g.  AGNs, SNe, GRBs), and in these rare instances, high
Strehl ratios are probably called for.

Accurate measurement with future space-based and ground-based imagers
is going to demand powerful software codes that take into account
the variable psf properties over the field.  The JWST psf will have
the 6-pointer structure arising from the Fourier transform of the
segmented mirrors. ELTs will have a similar psf structure, but
compounded by atmospheric distortion. One of the most serious of
these is anisoplanatism, i.e. field varying psf due to the slowly
varying field angle as the source tracks across the sky. It will
take complex computer codes to restore the photometric accuracy
(see the web site cfao.ucolick.org/meetings/psf\_reconstruction for
detailed discussions). The real gains in these codes are likely to
come from incorporating the extra information from the wavefront
sensor telemetry.

\smallskip\noindent
{\bf Configurable fields.} We mentioned above that the available
field at the focal plane of many telescopes is often much wider
than the exploited field of existing instruments. Detector real
estate, particularly IR arrays, is often a limiting cost in instrument
design, particularly so at IR and mid IR wavelengths.

But the imaged fields do not need to be defined by a contiguous
region (Bland-Hawthorn et al 2004). Indeed, for high redshift fields
(e.g. Hubble Ultra Deep Field [HUDF]), the information content can
be as low as 5\%, i.e. the fraction of pixels that contain useful
information.  Several observatories, in particular the AAO, are
working on robotic positioners that would allow random patches of
sky to be reformatted efficiently with relay optics to be packed
onto a wide-format detector. Such systems are likely to be a feature
of future instrument suites on ELTs, e.g.  deployable IFU wide-field
spectrographs (McGrath \& Haynes 2006), and multi-object AO systems
(MOAO; Hammer et al 2004).

\smallskip\noindent
{\bf Ultradeep imaging.}
Astronomers keenly await the awesome reach of the James Webb Space
Telescope (JWST) expected to launch in the middle of the next decade.
In a recent review, Gardner et al (2006) describe the scientific
potential of this facility.  The NIRCam imager will allow for diffraction
limited, broadband and intermediate band imaging (R$\sim$ 4, 10,
100) in the window $0.6-5\mu$m for a field of view and pixel sampling
comparable to the HST (2.2\arcmin$\times$4.4\arcmin) but at much
higher sensitivity, particularly in the mid IR. The MIRI camera
will allow broadband imaging in the window $5-27\mu$m over a
1.4\arcmin$\times$1.9\arcmin\ field.


On a similar timescale, we may expect to see results from one of
the proposed ground-based extremely large telescopes (ELT), in
particular, the 24.5m Giant Magellan Telescope (GMT), the 30m telescope
(TMT) or the recently announced European 42m ELT (E-ELT). These
telescopes will be optimized for near-infrared performance although
will extend into the optical and mid-IR. At the diffraction limit,
these telescopes will rival or even exceed the performance of the
JWST, particularly if the OH suppression problem is finally resolved
and the AO systems prove to be stable over arcmin-scale fields.  In
practice, there will be major gains from combining ground-based AO
imaging with space-based imaging of the same source (e.g. Vacca et
al 2007).

\begin{sidewaystable}
\centering
\begin{tabular}{llcllclllc}
instrument & mirror     &   D   & type    & window  &  AO   & field of view (\arcsec)                    & pixel scale (\arcsec)      & R             & N\&S \\
\\
GMOS       & Gemini-N   &   8   & Fibre   & Optical &  N    & 5$\times$7, 5$\times$3.5               & 0.2            & 1080$-$7100   & N \\
GMOS       & Gemini-S   &   8   & Fibre   & Optical &  N    & 5$\times$7, 5$\times$3.5               & 0.2            & 1080$-$7100   & Y \\
NIFS       & Gemini-N   &   8   & Slicer  & IR      &  AO   & 3$\times$3                          & 0.1            & 5000          & N \\
GNIRS      & Gemini-S   &   8   & Slicer  & IR      &  N    & 3.2$\times$4.8                      & 0.15           & 1700$-$5900   & N \\
ARGUS      & VLT        &   8   & Fibre   & Optical &  N    & 11.5$\times$7.3 or 6.6$\times$4.2     & 0.52, 0.3    & 19000$-$39000 & N \\
VIMOS      & VLT        &   8   & Fibre   & Optical &  N    & 54$\times$54 or 13$\times$13           & 0.67, 0.3    & 200$-$2500    & N \\
SINFONI    & VLT        &   8   & Slicer  & IR      &  AO,N & 8$\times$8, 3$\times$3, 0.8$\times$0.8   & 0.25$-$0.025 & 1500$-$4000   & N \\
IMACS-IFU  & Magellan   &  6.5  & Slicer  & Optical &  N    & 6.9$\times$5.0, 4.2$\times$5.0       & 0.2            & 1800$-$10,000 & N  \\
INTEGRAL   & WHT        &  4.2  & Fibre   & Optical &  N    & 7.8$\times$6.4, 33.6$\times$29.4       & 0.45, 2.70   & 200$-$10,000  & N \\
OASIS      & WHT        &  4.2  & Lenslet & Optical &  AO,N & 7.4$\times$10.3, 2.7$\times$3.7        & 0.26, 0.09   & 200$-$4000    & N \\
SAURON     & WHT        &  4.2  & Lenslet & Optical &  N    & 41$\times$33, 11$\times$9              & 0.94, 0.27   & 3000          & N \\
SPIRAL     & AAT        &  3.9  & Fibre   & Optical &  N    & 22$\times$11                        & 0.7            & 1500$-$13,000 & Y \\
SparsePak  & WIYN       &  3.5  & Fibre   & Optical &  N    & 72$\times$71(sparse)              & 4.7            & 5000$-$21,000 & N \\
DensePak   & WIYN       &  3.5  & Fibre   & Optical &  N    & 30$\times$45                        & 3              & 5000$-$21,000 & N \\
UIST       & UKIRT      &  3.8  & Slicer  & IR      &  N    & 3.3$\times$6.8                          & 0.24, 0.12   & 1000$-$4000   & N \\
PMAS       & Calar Alto &  3.5  & Fibre   & Optical &  N    & 8$\times$8                         & 0.5            & 1500$-$8000   & N \\
WIFES      & ANU        &  2.3  & Slicer  & Optical &  N    & 25$\times$31                        & 0.5            & 3000$-$7000   & N \\
CIRPASS    & Cambridge  &  $-$  & Fibre   & IR      &  AO,N & 13$\times$4.7, 9.3$\times$3.5          & 0.36, 0.25   & 3000          & N \\
\end{tabular}
\caption{A summary of operational integral field spectrographs. (1) instrument (2) telescope (3) telescope diameter in metres (4) technology (5) wavelength of operation (6) AO, natural seeing, or both (7) field of view in arcseconds (8) pixel scale in arcseconds (9) resolving power (10) nod \& shuffle operation?}
\end{sidewaystable}


\bigskip\noindent
{\bf\large 5. Imaging spectrographs}

\smallskip\noindent
{\bf Tunable imaging filters.}
I summarize the main technologies elsewhere (Bland-Hawthorn 2000).
The first general user tunable filter was the Taurus Tunable Filter
(TTF) that was operated at the AAT 3.9m and WHT 4.2m during the years
1995$-$2003. The scientific legacy is described in Bland-Hawthorn
\& Kedziora-Chudczer (2003).

There are several optical systems that are commissioned or close
to commissioning that are a direct consequence of the TTF. These
include the Maryland-Magellan Tunable Filter (MMTF), the Osiris
tunable filter on the GTC 10.2m (Cepa et al 2003), and the SALT
Tunable Filter. The Osiris spectrograph is particularly powerful
as it will provide tunable imaging (R=50$-$5000) and spectroscopic
capability over the range 370 to 1000 nm at high efficiency. The
JWST will also incorporate a restricted tunable filter for use at
infrared and mid-infrared wavelengths in the next decade.

Tunable filters are immensely powerful for conducting star formation
studies in both distant and nearby galaxies. The instruments are
tuned to narrow bands in order to provide a very high contrast
between emission lines and the neighbouring stellar continuum.
However, Ryder, Fenner \& Gibson (2005) demonstrate the power of
tunable filter imaging in tracing variations in stellar abundance
from absorption line variations over the face of galaxies. They
tuned the TTF bandpass to an equivalent Lick index, and control
bandpasses were observed to check the integrity of the measurement
over the field. Both methods are likely to be exploited extensively
with the new facilities.

\smallskip\noindent
{\bf Integral field and image slicer spectrographs.}
The main technologies are reviewed in van Breugel \& Bland-Hawthorn
(2000); a variety of science drivers is discussed by Cecil (2006),
Morris et al (2006) and Sharp et al (2004). In Table 1, I list the
main IFU spectrograph facilities in operation or soon to be realized.

As we have witnessed at this meeting, the SAURON project (TIGER
concept) has amply demonstrated the power of IFU spectroscopy for
stellar population studies (McDermid et al 2006; Kuntschner et al
2006).  Powerful general-user survey instruments like VIMOS on the
VLT and the GMOS IFU on Gemini have begun to deliver excellent data
on nearby galaxies.  Other facilities include INTEGRAL on the WHT
4.2m, SPIRAL at the AAT 3.9m and the WIFES image slicer at the ANU
2.3m (2008). All of these are used in natural seeing, offer resolving
powers of up to $\sim$few thousand, and have longest dimensions of
about $<$30\arcsec. The VLT GIRAFFE instrument offers smaller IFU
fields for multiple targets.

Table 1 shows the trend towards IFU spectrographs on 8m class
telescopes, but the power of the 4m class facilities should not be
underestimated. The 8m facilities give a fourfold increase in
collecting area but the pixel solid angle is an order of magnitude
smaller typically. This is not an obvious gain given that the surface
brightness of a galaxian stellar population is at or below the sky
level, although the gains can be substantial for cuspy core sources
or bright emission line sources.

Interesting recent developments are AO-assisted integral field
spectrographs (Rutten, Benn \& Mendez 2006), e.g. SINFONI at the
VLT and NIFS at Gemini South. While the fields of view are small
($64\times$32 format), the latest SINFONI observations of stellar
populations and kinematics in distant galaxies show the enormous
potential of these devices (Forster-Schreiber et al 2006).  An
exciting future facility is the MUSE IFU spectrograph slated for
the VLT in 2012 (Henault et al 2003). This will have an incredible
300$\times$300 format (0.2\arcsec\ pixels) offering comparable
spectral resolutions at optical wavelengths over an arcmin field.

\smallskip\noindent
{\bf Objective prism imaging.}  
With a dark IR sky background, old ideas will need to be revisited
(Bland-Hawthorn 2006). An interesting future prospect is the NIRspec
(1-5$\mu$m) and MIRI (5-10$\mu$m) objective prism spectroscopy on
JWST.  Both systems will offer R$\sim$100 spectroscopy over a
3.1\arcmin$\times$3.4\arcmin\ and 1.4\arcmin$\times$1.9\arcmin\
field respectively with a pixel sampling of about 0.1\arcsec.  In
the next section, I broach another concept to demonstrate how a
dark sky can be exploited in new ways.


\bigskip\noindent
{\bf\large 6. MAXIMUS -- a radical instrument concept}

On the back of recent developments in photonics, I now propose a
powerful new approach to imaging spectroscopy. The concept attempts
to {\tt maxim}ize the amount of {\tt us}eful spectral information
over a field while minimizing the number of spectral resolution
elements. It combines the power of imaging with the power of
spectroscopy. I give only a sketch at this time to illustrate how
a darkened night sky will allow us to explore new technological
avenues.

The instrument retains the constant spatial sampling (i.e.  matched
to the corrected seeing) across the field, but adapts the spectral
resolution to the average flux in a given pixel. This `prior
information' is supplied either by a snapshot image or a deep image
from another facility (e.g. HUDF)\footnote{One can envisage variants
where the spatial binning is varied over the field to conserve flux,
much like adaptive binning techniques at x-ray wavelengths.}.

MAXIMUS makes use of three ongoing developments in photonics: (i)
OH suppression through fibre Bragg gratings, (ii) integrated photonic
circuits where a fibre feeds light directly into a small ($\sim$few
cm) integrated photonic spectrograph (Bland-Hawthorn \& Horton
2006), (iii) photonic networks or switchyards.

With the night sky removed, one can now disperse individual pixels
at a resolving power that ensures the signal is fairly constant
over the observed data cube. Light from the telescope is reimaged
by a microlens array onto a bundle of optical fibres. We suppress
the night sky in each fibre photonically (Bland-Hawthorn 2006)
before directing individual fibres into an optical circuit with the
appropriate spectral resolving power. Light from individual fibres
is switched to the appropriate circuits via photonic networks much
like those already in use by the telecomm industry.

Consider a broadband image of a distant galaxy cluster with total
counts ranging from 10 to 10$^5$ counts with a background count
level of 10 counts, say. If all pixels are dispersed at R=4, the
skewed SNR distribution spans from 3 to about 320. If all pixels
are dispersed at R=4096, the spread is from 0.1 to 10. With the
MAXIMUS approach, the SNR variation is only a factor of 3, i.e.
from 3 to 10. This is a factor of 30 improvement in the use of
spectral resolution elements and allows for much longer exposures,
even in the presence of bright sources (e.g. guide stars).

For the data content of a typical astronomical image (e.g. resolved
stellar fields, galaxies), about 50\% of pixels are observed in
imaging mode (R$\sim$4), 45\% of pixels are dispersed at low
spectroscopic resolution (R$\sim$16, 64, 256), and about 5\% are
dispersed at intermediate and high resolution (R$\sim$1024, 4096).
There is a conservation principle at work here in that the required
photonic circuits at higher resolving power are larger and therefore
more expensive, but fewer are needed. In our example, the cores of
bright cluster galaxies and emission line sources will be resolved
at high resolution, the outer parts at medium resolution, and distant
faint blobs will have sufficiently low resolution to perform
photometric redshifts, for example. The dark sky will be detected
in single detector pixels in order to confirm that useful data have
not been overlooked.

The cost of an instrument can be measured in terms of the number
of resolution elements (spatial and spectral) that in turn relate
to the total number of detector pixels.  Assuming that the photonic
technologies are realized, this concept may provide the necessary
step to realizing the Million Element Integral Field spectrograph
(MEIFU) first proposed by the University of Durham (Content, Morris
\& Dubbeldam 2003).  Preliminary calculations suggest that, for the
same overall information content, MAXIMUS may require roughly 30
times fewer resolution elements than a traditional MEIFU design at
a fixed resolving power.


\bigskip\noindent
{\bf\large 7. Concluding remarks}

Over the next ten years, we can look forward to a time of deep,
wide-field imaging surveys from UV to infrared wavelengths, extending
to the mid infrared in the next decade. The time domain will identify
different populations through stellar variability.  There is a real
need for large surveys of galaxies observed in sub-arcsecond
conditions, at optical and infrared wavelengths. This will become
possible for the first time with the Pan-STARRS and LSST surveys
at optical wavelengths, and with the VISTA and UKIDSS surveys at
IR wavelengths. But beyond these surveys, there is a pressing need
for large surveys of galaxies observed at 0.2\arcsec\ FWHM or better,
particularly for resolved stellar work.  The HST has already
demonstrated the richness of information on these scales for Local
Group galaxies. In the best seeing conditions, star counts in the
outer parts of galaxies are achieving effective surface brightness
levels far below what is possible with diffuse light imaging.  With
the successful commissioning of near-IR MCAO systems, it is likely
that the community will look to push AO systems into the red optical
region. Since the number of actuators goes as a high power of the
photon energy, this will be challenging.

Beyond broadband imaging, the integral field spectrograph (and
variant) will continue to play a crucial role, particularly when
assisted by adaptive optics. These offer great advantages that we
have barely begun to explore. In the next decade, we will see these
used over wider fields of view, and with increased functionality
(e.g. field configurations). IFUs will be fundamental to ELT operation
for a host of reasons, e.g. correcting for deficiencies to high-order
atmospheric refraction that cannot be compensated by the field
corrector. There is also the exciting prospect of completely
suppressing the dominant OH signal such that an IFU can achieve
sensitivity on a par with the JWST at near-IR wavelengths (Bland-Hawthorn
2006).

Are there obvious technological arenas that we are not fully
exploiting?  There is not the space to discuss this question in
depth but there are a few that come to mind. Optical interferometry
has immense potential, both in terms of existing planned facilities
at the Keck and VLT, and proposed arrays that could challenge ELTs.
Another area is multiband simultaneity, i.e. observing the same
patch of sky with wide-field detectors in multiple bands simultaneously,
a concept that is exploited by digital cameras.  The increased
miniaturization of integrated circuitry means that we can cram more
logic into each detector pixel (e.g. active pixel sensing). Tonry
\& Luppino (2000) raise the prospect of an arrty of microprisms
that disperse light over complex pixel structures.  One can envisage
equivalent technologies, i.e. a cascade of beam splitters that
divide up the spectrum and redirect the light to different wide-field
detectors.  An alternative strategy is to use layered materials
that have increasing opacity to the incoming photons with material
depth.

But for now, the largest gains are unlikely to come from advances
in detector technology. The community awaits wide-field AO systems
on the 8-10m ground-based telescope, the refurbishment of HST with
WFC3, and four large survey telescopes on sub-arcsecond sites.
Beyond here, there is the prospect of JWST and one or more ELTs, all
of which offer extraordinary gains over existing systems.  There
are enormous technological challenges to be overcome, not least the
problem of realizing stable and accurate AO-assisted photometry
with extremely large telescopes.

\begin{acknowledgments}
I am indebted to S.D. Ryder for comments on this manuscript, and to 
R.G. Sharp (AAO) for preparing Table 1 and for his insights on IFU 
spectroscopy. 
\end{acknowledgments}

\end{document}